# Comparative study of $Mo_2Ga_2C$ with superconducting MAX phase $Mo_2GaC$: a first-principles calculations

## M. A. Ali[1,*], M. R. Khatun[2], N. Jahan[1], M. M. Hossain[1]

[1]Department of Physics, Chittagong University of Engineering and Technology, Chittagong-4349, Bangladesh.
[2]Department of Physics, University of Rajshahi, Rajshahi-6205, Bangladesh.
[*]Corresponding author. E-mail:ashrafphy31@cuet.ac.bd

**Abstract**

The structural, electronic, optical and thermodynamic properties of $Mo_2Ga_2C$ are investigated using density functional theory (DFT) within the generalized gradient approximation (GGA). The optimized crystal structure is obtained and the lattice parameters are compared with available experimental data. The electronic density of states (DOS) is calculated and analyzed. The metallic behavior for the compound is confirmed and the value of DOS at Fermi level is 4.2 states per unit cell per eV. Technologically important optical parameters (e.g., dielectric function, refractive index, absorption coefficient, photo conductivity, reflectivity, and loss function) have been calculated for the first time. The study of dielectric constant ($\varepsilon_1$) indicates the Drude-like behavior. The absorption and conductivity spectra suggest that the compound is metallic. The reflectance spectrum shows that this compound has the potential to be used as a solar reflector. The thermodynamic properties such as the temperature and pressure dependent bulk modulus, Debye temperature, specific heats, and thermal expansion coefficient of $Mo_2Ga_2C$ MAX phase are derived from the quasi-harmonic Debye model with phononic effect also for the first time. Analysis of $T_c$ expression using available parameter values (DOS, Debye temperature, atomic mass etc.) suggests that the compound is less likely to be superconductor.

## 1. Introduction

Recently, scientific community have paid their significant attention to an unusual class of layered ternary carbides and nitrides, the so called MAX phases because of their outstanding combination of properties, some of which are like ceramics and the others metallic [1, 2]. To be specific, the properties for which the metal are applicable in industrial scale are the machinability, damage tolarance, thermal and electrical conductivity, are possessed by these materials. They also possess the properties of ceramics such as high elastic stiffness, refractory nature, and resistant to high-temperature oxidation [3]. The outstanding combination of these properties makes them attractive for potential applications in diverse fields from defense materials to electronic devices such in defense, aerospace, automobile, medical, nuclear reactor and portable electronic devices where they are already being used. The charismatic uniqueness of the MAX phases is motivating numerous researches to look forward that they can open the way to practical commercial applications for these materials in future. So far, more than 70 different MAX phases have been experimentally synthesized [4] and also a good number of MAX phases have been theoretically predicted. The research in searching the new MAX phases is growing fast in order to discover many more new MAX phase compounds due their properties mentioned above. The $M_2AX$ (211) phases including solid solutions with M = Ti, V, Cr, Nb, Ta, Zr, Hf; A= Al, S, Sn, As, In, Ga, and X = N, C, have been studied extensively both experimentally and theoretically [5 and the references (1-29) therein, 6-11]. Among them $Mo_2GaC$ [12] is one the important MAX phase showing superconducting characteristics with $T_c$ ~ 4.0 K. Very recently, Hu et al. [13] reported on the discovery of a totally new ternary hexagonal $Mo_2Ga_2C$ phase, is a counterpart of superconducting $Mo_2GaC$, assumed to be the first member of a distinct large family closely related to the MAX phases.

Studies of $Mo_2Ga_2C$ phase have been reported in the literature. The structural and compositional analysis has been addressed by Lai et al [14]. Elastic and electronic properties have been investigated by M. A. Hadi [15]. Another plausible metastable structure with close-packed Ga layers is predicted from density functional calculations by Wang et al [16]. However, though structural, elastic and electronic properties are studied but the thermodynamic and optical properties were not taken into account. Moreover, due to the similarity of structure and electronic bonding with superconducting $Mo_2GaC$, $Mo_2Ga_2C$ might also be superconductor. The possibility of this property is not taken in consideration in the previous studies.

The thermodynamic properties are very important in solid states science and consider as the basis for industrial application of solids because material's behaviour can be obtained from thermodynamic properties under high temperatures as well as high pressure. Moreover, the optical properties provide the information about the electronic response of the materials which are related to the electronic properties of solids [17]. Therefore, an investigation of these properties is significantly necessary for fundamental physics and potential applications.

In this work, we have aimed to provide some additional information to the existing data on the physical properties of $Mo_2Ga_2C$ phase by using the first-principles method, and we have especially focused on the possible occurrence of superconductivity, thermodynamics and optical properties.

## 2. Computational Methods

The calculations were carried out using the CAmbridge Serial Total Energy Package (CASTEP) code [18] based on the density-functional theory [19]. The generalized gradient approximation (GGA) of Perdew-Burke-Ernzerhof(PBE) scheme [20] is used as the exchange and correlation function. The electrostatic interaction between valence electron and ionic core is represented by the ultrasoft pseudopotentials, and the cutoff energy for the plane wave expansion is 550 eV. A $17 \times 17 \times 3$ k-point mesh of Monkhorst-Pack [21] scheme was used for integration over the first Brillouin zone. The Broyden–Fletcher–Goldfarb–Shanno (BFGS) algorithm [22] techniqueis was applied to optimize the atomic configuration and density mixing is used to optimize the electronic structure.

## 3. Results and discussion

### 3.1. Structural properties

Similar to the all other MAX phases, the new compound $Mo_2Ga_2C$ crystallizes in the hexagonal system with space group $P6_3/mmc$ (No. 194). Not only similar in crystal system but also similar in unit cell structure with 211 MAX phases such as the unit cell contains two formula units. The number of atoms per unit cell is not same. In the case of 211 MAX phases there are 8 atoms in their unit cell, whereas the new compound $Mo_2Ga_2C$ has 10 atoms in its unit cell. There is a difference at the position of Ga atoms of $Mo_2Ga_2C$ (4f Wyckoff position) and $Mo_2GaC$ (2d Wyckoff position).The lattice constant *a* remains almost unchanged, but due to the extra Ga layer along *z*-axis, lattice constant *c* is changed (Table 1). Fig. 1 Show the unit cell structure of $Mo_2Ga_2C$ and $Mo_2GaC$. Table shows the lattice constant of $Mo_2Ga_2C$ along with experimental data. Table 1 also contains the lattice constant of $Mo_2GaC$ for comparison. The calculated lattice parameters are in reasonable agreement with the experimental results.

Table 1 Lattice constants and atomic fractional coordinates of $Mo_2Ga_2C$ and $Mo_2GaC$.

| Phase | $a$(Å) | $c$(Å) | Ref. | Atomic position | | |
|---|---|---|---|---|---|---|
| | | | | Mo | Ga | C |
| $Mo_2Ga_2C$ | 3.047 | 18.164 | This Study | 4f | 4f | 2a |
| | 3.033 | 18.081 | Expt. [13] | | | |
| | 3.031 | 18.11 | Expt. [14] | | | |
| | 3.05374 | 18.13445 | Theo. [15] | | | |
| $Mo_2GaC$ | 3.064 | 13.178 | This Study | 4f | 2d | 2a |
| | 3.01 | 13.18 | Expt. [2] | | | |

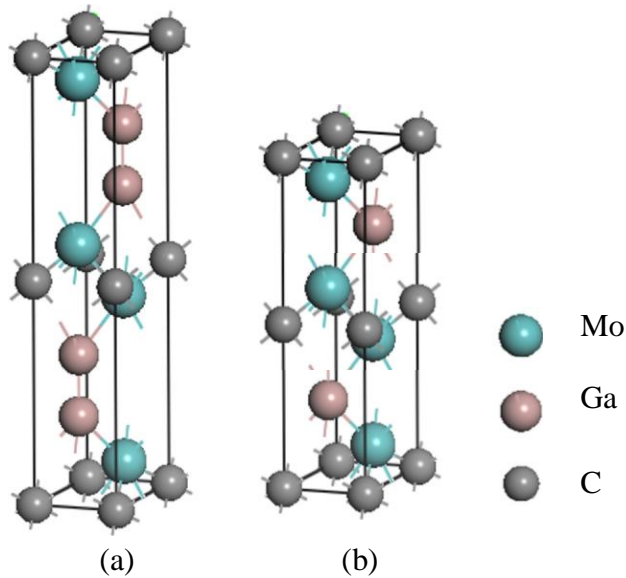

(a)      (b)

Fig. 1. The unit cell structure of (a) $Mo_2Ga_2C$ and (b) $Mo_2GaC$.

*3.2 Electrons DOS: Possibility of Superconductivity*

Fig. 2. Shows the total and partial density of states (DOS) of $Mo_2Ga_2C$ (left) and $Mo_2GaC$ (right). We have used the DOS values to predict the possibility of superconductivity in $Mo_2Ga_2C$ compared with $Mo_2GaC$. We have also predicted the possibility of superconductivity in some materials compared with iso-structural superconducting phase in the same way [23]. The calculated DOS at the Fermi level for $Mo_2Ga_2C$ is 4.2 states per unit cell per eV, whereas for $Mo_2GaC$ it is found to be 4.5 states per unit cell per eV. It is significant to explain the nature of electrons close to the Fermi surface because these electrons will contribute to form the superconducting state of materials. It is found that the DOS at the Fermi level originates mainly from Mo 5d states. In order to make clear the possible occurrence of superconductivity in $Mo_2Ga_2C$, investigation about electron-phonon coupling properties of the compound should be performed. Electron-phonon coupling can be expressed as $\lambda = N(E_F)V$, where $V$ is the degree of the inter-electron attractive interaction. Again, McMillan's formula for $T_c$ [24] is proportional to Debye temperature ($\theta_D$) and an exponential term involving electron-phonon coupling constant, $\lambda$

= $N(E_F)<I^2>/M<\omega^2>$. Here M is the relevant atomic mass, $<I^2>$ the square of the e-phonon matrix element averaged over the Fermi surface, $<\omega^2>$ is the relevant phonon frequency squared, $N(E_F)$ is the DOS at the $E_F$. This may be helpful to understand the superconductivity of $Mo_2Ga_2C$. It can be seen from the expression that DOS would affect $T_c$ only if the $<I^2>$ is the same for $Mo_2Ga_2C$ and $Mo_2GaC$. The calculated values of $\theta_D$ are 471.2 and 483.8 K for $Mo_2Ga_2C$ and $Mo_2GaC$ [15], respectively. The $T_c$ equation when analyzed with all these factors of $Mo_2Ga_2C$ and compared with superconducting $Mo_2GaC$ indicates that the $Mo_2Ga_2C$ compound is less likely to be superconductor. If the superconductivity in $Mo_2Ga_2C$ will be confirmed in future then expected $T_c$ value will also be very close that of $Mo_2GaC$ because $T_c$ for $Mo_2Ga_2C$ and $Mo_2GaC$ may be connected to the phonon system. Indeed, as for $Mo_2Ga_2C$ and $Mo_2GaC$, the lattice parameter $a$, which defines the intra-atomic distances inside the conducting blocks, remains unchanged. This can lead to the same coupling constant $\lambda$. Definitely, these are assumptions, to be sure about these assumptions the phonon spectra should be calculated. Finally, we note, the superconductivity for $Mo_2Ga_2C$ has been not reported yet, the factors discussed here compared with superconducting $Mo_2GaC$ ($T_C \sim 4$ K) allow us to assume the emergence of low-temperature superconductivity for $Mo_2Ga_2C$, and we believe that related experiments will be of high interest.

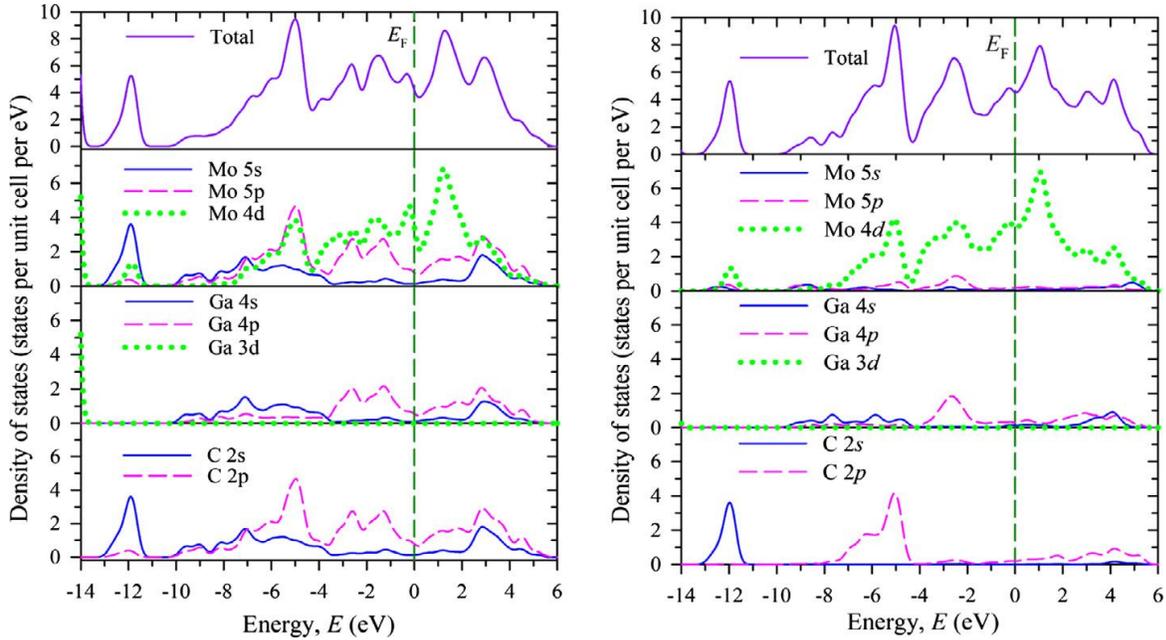

Fig. 2. Total and partial density of states (DOS) of $Mo_2Ga_2C$ (left) and $Mo_2GaC$ (right).

*3.4. Optical properties*

Optical properties are used to describe the materials behavior when electromagnetic radiation is incident on the materials. In order to describe the response of $Mo_2Ga_2C$ to electromagnetic radiation we have calculated some important optical constants for the first time. The methods by which the optical constants are calculated can be found elsewhere [23, 25].

The optical constants of $Mo_2Ga_2C$ for (100) polarization direction are shown in Fig. 3 (left and right panel). To smears out the Fermi level for effective k-points on the Fermi surface, we used a 0.5 eV Gaussian smearing. When light of sufficient energy incident onto a material, it causes electrons transition from valence to conduction band. This electrons transition takes part to contribute to the optical properties of metal-like systems which affects mainly the low energy infrared part of the spectra. To calculate the dielectric function of metallic $Mo_2Ga_2C$, a Drude term with unscreened plasma frequency 3 eV and damping 0.05 eV has been used.

The imaginary part, $\varepsilon_2(\omega)$ of the dielectric function, $\varepsilon(\omega)$ dominated the electronic properties of crystalline material, which depicts the probability of photon absorption. The peaks of $\varepsilon_2(\omega)$ are associated with electron excitation. There is only one prominent peak around 2.0 eV [Fig. 3 (b, left panel)]. The large negative value of $\varepsilon_1$ is also observed in Fig. 3 [Fig. 3 (a, left panel)] which is an indication of Drude-like behavior of metals. The refractive index is another technically important parameter for optical materials for its technological applications in optical devices. The spectra for refractive index $n$ is demonstrated in Fig. 3 (a, right panel). The static refractive index $n(0)$ is found to have the value of ~ 7 for $Mo_2Ga_2C$ while this value is 17.53 for $Mo_2GaC$ [26].

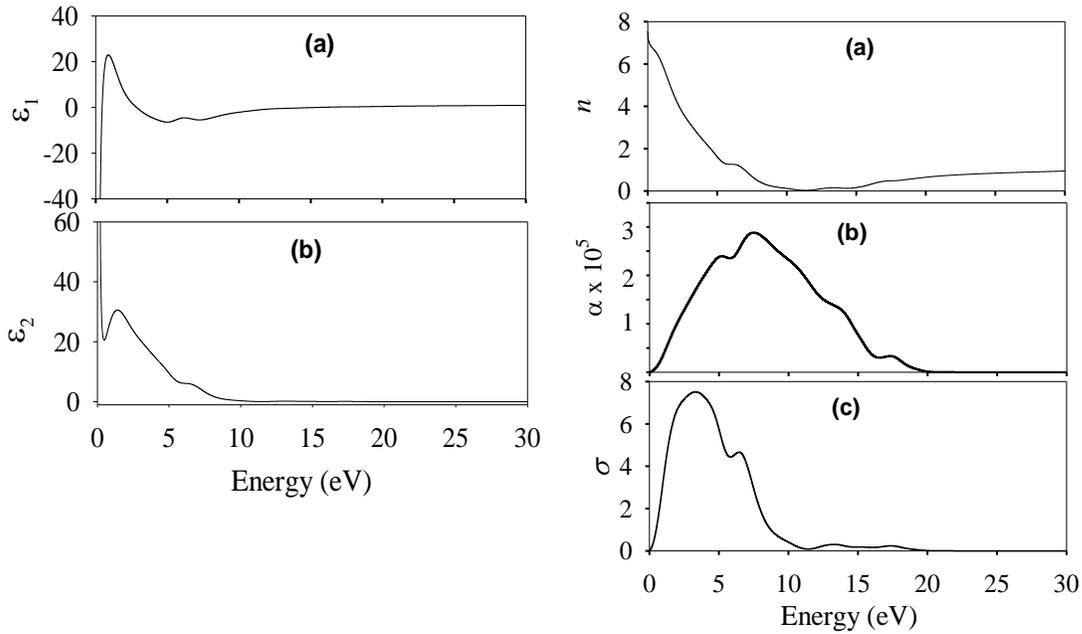

Fig. 3. (a) Real part & (b) imaginary part of dielectric constant, $\varepsilon_1$ and $\varepsilon_2$, respectively (left panel) and (a) refractive index, $n$ (b) absorption coefficients, $\alpha$ (c) conductivity, $\sigma$ (right panel) of $Mo_2Ga_2C$.

Fig.3 (b, right panel) shows the absorption coefficient spectra of $Mo_2Ga_2C$ which reveal the metallic nature of the compound since the spectra starts with non-zero value. Interestingly, a strong absorption coefficient is observed in the UV range. Moreover, the absorption coefficient is weak in the IR and is continuously increasing towards the UV region, reaches a maximum value at 7.7 eV. Based on these results, $Mo_2Ga_2C$ is a promising absorbing material in UV region. Since the materials with high absorption coefficients indicates the absorption of photons is increased in the materials, which excite electrons from the valence band to the conduction band. These materials are very important for optical and optoelectronic devices in the visible and ultraviolet energy regions. The band structure of the materials shows no band gap, which indicate that the photoconductivity starts at zero photon energies as shown in Fig. 3 (c, right panel). This type of photoconductivity confirms the good metallic nature of this compound. The reported absorption coefficient and photoconductivity spectra of $Mo_2GaC$ are almost same [26].

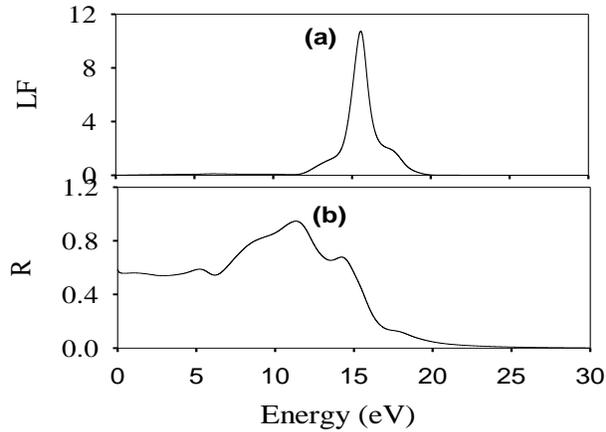

Fig. 4 (a) Loss Function and (b) Reflectivity of $Mo_2Ga_2C$.

The loss function $L(\omega)$, defined as energy loss of an electron with high velocity passing through the materials, shown in Fig. 3 (right panel). In addition, the peaks in the $L(\omega)$ spectrum represent a plasma resonance property (a collective oscillation of the valence electrons). In our present case, the energy loss function curve is characterized by a peak which is known as bulk plasma frequency $\omega_P$, occurs at $\varepsilon_2 < 1$ and $\varepsilon_1 = 0$. In Fig. 4, the value of the effective plasma frequency $\omega_P$ is found to be ~ 16 eV which lower than [17.2 eV] that of $Mo_2GaC$ [26]. If the frequency of incident photon is greater than $\omega_P$, then the material becomes transparent.

The reflectivity curve is also shown in Fig. 4 (b). It is found that the reflectivity curve starts with a value of ~ 0.58 and exhibit no significant changes in the energy range up to ~ 6.0 eV, rises to a maximum value of ~ 0.9 at ~ 12 eV. $Mo_2Ga_2C$ has roughly the similar reflectivity spectra as those by other 211 and/or other MAX phases [26]. The value of reflectivity remains always above 44%. Le et al [27] reported $Ti_3SiC_2$, having the average reflectivity ~ 44% in the visible light region, as a nonselective characteristics which is responsible for solar heat reducing. Moreover, the reflectivity spectrum is steady and stable in wide range (~ 6.0 eV) and then increases gradually. Therefore it is expected that $Mo_2Ga_2C$ compound is also appealing for the practical usefulness as a coating on spacecrafts to avoid solar heating. Moreover, the peak of loss function is associated with the trailing edges of the reflection spectra. For example, the peak in $L(\omega)$ occurring at ~ 16 eV corresponds to an abrupt decrease in reflectivity.

### 3.4 Thermodynamics properties

The study of thermodynamic properties of materials permits a deeper understanding on the specific behavior of materials under high temperature and pressure. The thermodynamic properties of $Mo_2Ga_2C$ have been investigated using quasi-harmonic Debye approximation [28, 29]. The data calculated using this method are in good agreement with experimental data proved by several authors [30, 31]. The temperature (0-1000 K) and pressure (0-50 GPa) dependent polycrystalline aggregate properties including bulk modulus, Debye temperature, specific heats and thermal expansion coefficients of $Mo_2Ga_2C$ have been calculated for the first time. The volume and total energy of $Mo_2Ga_2C$ calculated by the methodology described section 2, were used as input data in Gibbs program. The methodology by which the volume and total energy used as input in Gibb's program can be found in elsewhere [32].

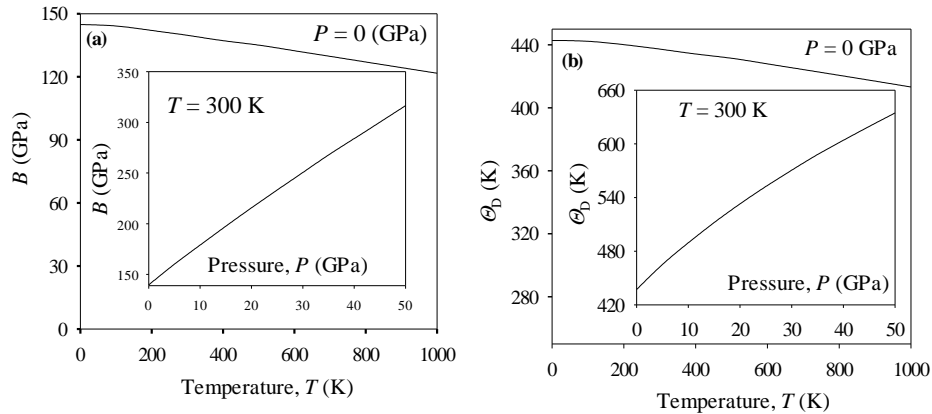

Fig. 5. Temperature dependence of (a) bulk modulus, $B$ and (b) Debye temperature, $\Theta_D$ of $Mo_2Ga_2C$. *Insets* show pressure dependence.

The bulk modulus, $B$ at 0 GPa of $Mo_2Ga_2C$ as function of temperature is shown in Fig. 5(a); the inset represents $B$ as a function of pressure. At ambient condition, the $B$ of $Mo_2Ga_2C$ is lower than that of $Mo_2GaC$ [15]. It can be found from figure that the value of $B$ are nearly flat for temperature 0 to 100 K. Above 100 K, $B$ decreases slowly in a slightly nonlinear way up to 1000 K. It is here noted that $B$ reduces by ~ 5% from 0 K to 1000 K. The value of $B$ at room temperature is shown in inset as a function of pressure. It is observed that $B$ increases with increasing pressure at a given temperature and decreases with increasing temperature at a given pressure, because the effect of increasing pressure on material is similar as decreasing temperature of material, which means that the increase of temperature of the material causes a reduction of its hardness. This phenomenon is well consistent with the trend of volume of the material although it is not shown in figure.

Fig. 5(b) displays the temperature dependence of Debye temperature, $\Theta_D$ of $Mo_2Ga_2C$ at $P = 0$ GPa. The *inset* of the figure shows $\Theta_D$ as a function of pressure at room temperature. The Debye temperature $Mo_2Ga_2C$ is also lower than that of $Mo_2GaC$ at ambient condition [15]. At fixed pressure, $\Theta_D$ decreases with increasing temperatures and at fixed temperature it increases with increasing pressure. These results indicate the change of the vibration frequency of particles with pressure and temperature. Most other solids have weaker bonds and far lower $\Theta_D$; consequently, their heat capacities have almost reached the classical Dulong–Petit value of $3R$ at room temperature as can be seen from Fig. 6(a). If it seems that the harder is the solid, the higher is the $\Theta_D$, and the slower is the solid to reach its classical $C_V$ of $3R$, this is not a coincidence.

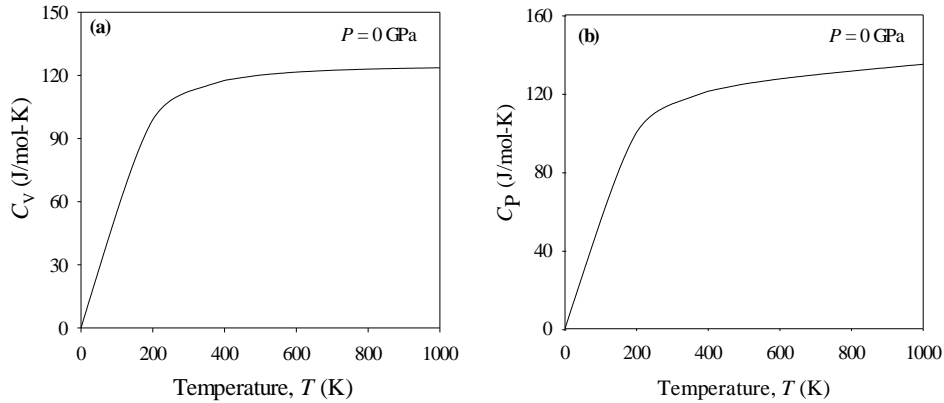

Fig. 6. Temperature dependence of lattice specific heat at constant volume (a) and specific heat at constant pressure (b) of $Mo_2Ga_2C$

The lattice heat capacity of a substance is a measure of how well the substance stores heat. The temperature dependence of $C_V$ is governed by the details of vibrations of the atoms and could be determined from experiments. It is worthwhile to outline that the Debye model correctly predicts the low-temperature dependence of the heat capacity at constant volume, which is proportional to $T^3$ [33]. It also recovers the Dulong–Petit law at high temperatures [34]. The heat capacities at constant-volume ($C_V$) and constant-pressure ($C_P$) of $Mo_2Ga_2C$ as a function of temperature are displayed in Fig. 6 (a, b). The temperature is limited to 1000 K to reduce the possible effect of anharmonicity. The heat capacities, $C_V$ and $C_P$ both increase with the increase of applied temperature due to the phonon thermal softening occurs as the temperature is increased. The difference between $C_P$ and $C_V$ for the phase is calculated by $C_P - C_V = \alpha_V^2(T)BTV$, where $\alpha_V$ is the volume thermal expansion coefficient. The difference of heat capacities is very small, which is due to the thermal expansion caused by anharmonicity effects. It is shown in Fig. 6 that the heat capacities of $Mo_2Ga_2C$ increase quickly with increasing temperature at low temperature range ($T < 300$ K) and thereafter rises slowly up to 700 K and finally approaches a saturation value. However, in the low temperature region, the heat capacities follow the Debye T3 power-law whereas at high temperature limit, these approach the Dulong-Petit limit of $C_V = 3nNk_B =$

120 J/mol- K. These results reveal that the interactions between ions in $Mo_2Ga_2C$ have great effect on heat capacities especially at low $T$.

The volume thermal expansion coefficient, $\alpha_V$ as a function of temperature and pressure are shown in Fig. 7. It is also found that $\alpha_V$ increases rapidly with increasing temperature at low temperature region of $T < 300$ K and increases gradually after 300 K. The calculated value of $\alpha_V$ at 300 K is $3.6\times10^5$ $K^{-1}$ which is greater than that of $Mo_2GaC$ [26] due to the lower bulk modulus value of $Mo_2Ga_2C$ than $Mo_2GaC$. It is established that the volume thermal expansion coefficient is inversely related to the bulk modulus of a material. The estimated linear expansion coefficients ($\alpha = \alpha_V/3$) is $1.2 \times 10^{-5}$ $K^{-1}$. It can also be found that $\alpha_V$ decreases gradually with increasing pressure.

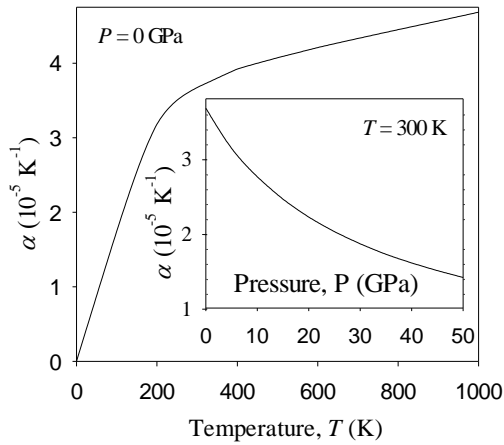

Fig. 7 The volume thermal expansion coefficient of $Mo_2Ga_2C$ as a function of temperature. The inset shows the pressure dependence.

## 4. Conclusions

A first time investigation of optical and thermodynamic properties of $Mo_2Ga_2C$ and the prediction of the occurrence of superconductivity is carried out by means of first principles method. The thermodynamic properties are derived from the quasi-harmonic Debye model with phononic effect. The energy bands around from the Fermi level are mainly from Mo $4d$ states, suggesting that the Mo $4d$ states dominate the conductivity. The analysis of the electronic band structure indicates the metallic behavior of the compound, which is also confirmed by the study of absorption and conductivity spectra. All optical functions are calculated in polarization direction (100) and analyzed in details. The results are in good agreement with other reported results of $Mo_2Ga_2C$. The results are also compared with that the $Mo_2GaC$ where available. Based on our present study, it is worth to say that $Mo_2Ga_2C$ material could be used as technologically important material.